\documentclass[prb,twocolumn]{revtex4}   % Physical Review B
\usepackage{graphicx}

\begin{document}

%%%%%%%%%%%%%%%%%%%%%%%%%%%%%%%%%%%%%%%%%%%%%%%%%%%%%%%%%%%%%%%%%%%%%%%
\title
{Frustrated Heisenberg antiferromagnet on a pyrochlore lattice}
%%%%%%%%%%%%%%%%%%%%%%%%%%%%%%%%%%%%%%%%%%%%%%%%%%%%%%%%%%%%%%%%%%%%%%%

\author
{ 
Akihisa Koga and Norio Kawakami
}

\affiliation
{
Department of Applied Physics, Osaka University, Suita, Osaka 565-0871, Japan
}

\date{\today}

\begin{abstract}
We investigate quantum phase transitions for the
$s=1/2$ antiferromagnetic Heisenberg model on a pyrochlore lattice.
By means of a series expansion starting from isolated tetrahedra,
the ground-state phase diagram is determined.
When the ratio of the two competing exchange couplings is varied,
the first-order (second-order) quantum phase transition occurs between
the two spin gap phases (the spin-gap and the antiferromagnetic phases).
We also discuss some properties expected for 
the $s=1$ pyrochlore spin system.
\end{abstract}

\pacs{Valid PACS appear here}% 

\maketitle
%%%%%%%%%%%%%%%%%%%%%%%%%%%%%%%%%%%
%\section{Introduction}
%%%%%%%%%%%%%%%%%%%%%%%%%%%%%%%%%%%
Geometrically frustrated magnetic materials
have been the subject of considerable interest recently.
A typical example is a class of transition-metal oxides
with pyrochlore structure. In particular, it was
reported that the specific-heat coefficient  is 
exceptionally large for a metallic 
compound  $\rm LiV_2O_4$.\cite{LiV2O4} 
It was recently pointed out that the frustration caused by a tetrahedral 
network of vanadium ions may
be important to understand the heavy-fermion behavior in 
this compound, \cite{Kaps} 
stimulating further intensive studies on the related 
systems with pyrochlore structure.
Such frustration effects should be much more prominent for
 quantum spin systems.
Theoretical studies \cite{Harris,Canals,Isoda}
for the $s=1/2$ quantum spin model
on a pyrochlore lattice were first done by Harris et al.,\cite{Harris} who
pointed out the possibility of the dimerized ground state
by exploiting a field theoretic approach.
Canals and Lacroix\cite{Canals} clarified that the ground state of the model 
is a spin-liquid state with the spin gap.
They found that the neutron diffraction cross section\cite{Y(Sc)Mn2} observed 
in $\rm Y(Sc)Mn_2$ is in fairy good agreement 
with their results.\cite{Canals}
Isoda and Mori,\cite{Isoda} however, used a bond-operator approach to
suggest that the ground state may be described by a
RVB-like tetrahedral (plaquette) singlet state, 
which is different from the dimer-singlet state known so far.
Furthermore, the possibility of the "topological spin glass" was
pointed out in $\rm Y_2Mo_2O_7$,\cite{Y2Mo2O7} making
this issue more attractive and challenging.

\begin{figure}[htb]
\begin{center}
\includegraphics[height=4cm]{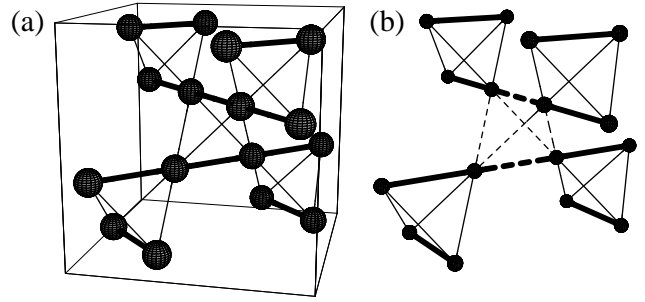}
\end{center}
\caption{(a) Frustrated antiferromagnetic spin model on a pyrochlore lattice.
Bold and thin-solid lines represent 
the exchange couplings $J$ and $J'$, respectively.
(b) Initial configuration for the series expansion:
Broken lines represent the perturbed bonds, $\lambda J$ and $\lambda J'$ 
(see text).
}
\label{fig:pyrochlore}
\end{figure}

In this paper, we investigate the $s=1/2$ quantum spin model 
on a pyrochlore lattice with competing antiferromagnetic 
interactions shown in Fig. \ref{fig:pyrochlore} (a).
Our system may describe some pyrochlore-lattice compounds
such as 
$\rm Y(Sc)Mn_2$\cite{Y(Sc)Mn2} as well as
$\rm GeCu_2O_4$\cite{GeCu2O4} found recently.
We study the ground-state phase diagram 
and clarify the role of the geometrical frustration
by studying quantum phase transitions 
by means of series expansion techniques.\cite{series}
We also discuss the $s=1$ case briefly.

%%%%%%%%%%%%%%%%%%%%%%%%%%%%%%%%%%%%%%%%%%%%%%%%%
%\section{series expansion approach}
%%%%%%%%%%%%%%%%%%%%%%%%%%%%%%%%%%%%%%%%%%%%%%%%%
Let us first consider the $s=1/2$  spin model
on a pyrochlore lattice, which is described by the following Hamiltonian
\begin{eqnarray}
H=J\sum_{(i,j)}{\bf S}_i\cdot{\bf S}_j+J'\sum_{<i,j>}{\bf S}_i\cdot{\bf S}_j,
\label{eq:H}
\end{eqnarray}
where $(i,j)$ denotes a pair of two adjacent sites connected by the 
thick bond in Fig. \ref{fig:pyrochlore} (a), whereas $<i,j>$ 
is that for the thin bond. Both of the exchange couplings
$J$ and $J'$  are assumed to be antiferromagnetic. 
%%%%%%%%%%%%%%%%%%%%%%%%%%%%%%%%%%%%%%%%%%%%%%%%%%%%%%%%%%%%%%%%%
%%%%%%%%%%%%%%%%%%%%%%%%%%%%%%%%%%%%%%%%%%%%%%%%%%%%%%%%%%%%%%%%%
For the compound $\rm Y(Sc)Mn_2$, 
we may take $J/J' \sim 1$,\cite{Y(Sc)Mn2}  while for 
$\rm GeCu_2O_4$, it is to be $J/J' \sim 6$.\cite{GeCu2O4}

Before proceeding with the analysis, we 
 note that, in some limiting cases, the spin system 
is reduced to simple models whose nature can be easily understood.
For $J'=0$, it is equivalent to the $s=1/2$ massless Heisenberg spin chain,
whereas an antiferromagnetically ordered state is stabilized  for $J=0$
since the model has a three-dimensional structure 
without frustration in this case. To determine the phase diagram,
we study what kind of quantum phase transition occurs
when the competing interactions are varied.  For this purpose,
we use the series expansion method, \cite{series}
which  has an advantage to deal with frustrated spin systems 
in higher dimensions.
In fact, it was successfully applied to frustrated spin systems 
such as $J_1-J_2$ model,\cite{J1J2} plaquette system,\cite{plaquette} 
orthogonal-dimer system.\cite{dimer}
To apply the series expansion method to the pyrochlore-lattice system,
we first divide the original Hamiltonian eq. (\ref{eq:H}) into two parts as
$H=H_0+\lambda H_1$ by introducing an auxiliary parameter $\lambda$,
 where $H_0 (H_1)$ represents the unperturbed (perturbed) 
Hamiltonian. Note that the system is reduced to the original
model for $\lambda=1$.
We here choose a tetrahedron composed of four spins 
as a starting configuration ($H_0$),\cite{Harris,Canals}
and then connect each tetrahedron 
via tetrahedral bonds labeled by $\lambda J$ and $\lambda J'$ 
[see Fig. \ref{fig:pyrochlore} (b)].
The Hamiltonian $h$ for an isolated tetrahedron in $H_0$ is given by 
%%%%%%%%%%%%%%%
\begin{eqnarray}
h&=&J\left({\bf S}_1\cdot{\bf S}_2+{\bf S}_3\cdot{\bf S}_4\right)+
J'\left({\bf S}_1+{\bf S}_2\right)\cdot\left({\bf S}_3+{\bf S}_4\right).
\label{eq:iso}
\end{eqnarray}
%%%%%%%%%%%%%%%%%%
The energy eigenvalues, $E$, of the tetrahedron for a given  $j=J/J'$
are listed in Table \ref{I},
where $S_{12} (S_{34})$ represents the combined spin 
$S_1+S_2 (S_3+S_4)$, and $S_{\rm total}$ is the total spin. 
%%%%%%%%%%%%%%%%%%%%%%%%%%%%%%%%%%%%%%%%%%%%%%%%%%%%%%%%%%%%%%%%%%%%%%%%%
\begin{table}[htb]
\caption{Eigenvalues of the $s=1/2$ spin system on 
an isolated tetrahedron.}\label{I}
%\begin{ruledtabular}
\begin{tabular}{cccc|cccc}
\toprule 
$S_{12}$      &0&1&0&\multicolumn{3}{c}{1}\\
$S_{34}$      &0&0&1&\multicolumn{3}{c}{1}\\
$S_{\rm total}$&0&1&1&0&1&2\\
\colrule
$E/J'$&$-\frac{3}{2}j$&$-\frac{1}{2}j$&$-\frac{1}{2}j$&
$-2+\frac{1}{2}j$&
$-1+\frac{1}{2}j$&$1+\frac{1}{2}j$\\
\botrule
\end{tabular}
%\end{ruledtabular}
\end{table}
%%%%%%%%%%%%%%%%%%%%%%%%%%%%%%%%%%%%%%%%%%%%%%%%%%%%%%%%%%%%%%%%%%%%%%%%%
It is seen in this table that for $0<j<1$, 
the isolated tetrahedron has a plaquette-singlet ground state
with $S_{12}=S_{34}=1$ and $S_{\rm total}=0$. 
On the other hand, for $j>1$  we have
the dimer ground state with $S_{12}=S_{34}=S_{\rm total}=0$.
The phases specified by these singlets are referred to 
as the plaquette and dimer phases, respectively.

\begin{figure}[htb]
\begin{center}
\includegraphics[width=8cm]{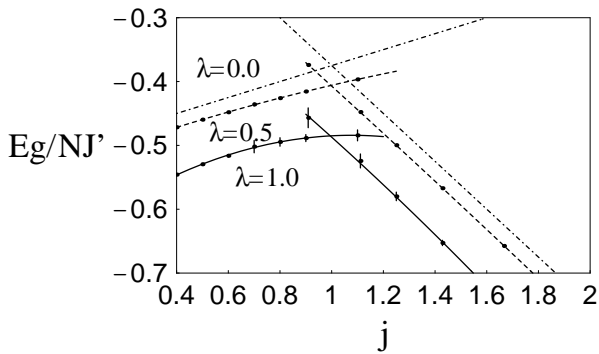}
\end{center}
\caption{Ground state energy for the dimer (right-side) 
and the plaquette (left-side) phases
for various values of $\lambda$.
The energy for $\lambda=0.0, 0.5$ and $1.0$ is shown as the dot-dashed, 
dashed and solid lines, respectively.
}
\label{fig:eg}
\end{figure}

%%%%%%%%%%%%%%%%%%%%%%%%%%%%%%%%%%%%%%%%%
%\subsection{First-order transition}
%%%%%%%%%%%%%%%%%%%%%%%%%%%%%%%%%%%%%%%%%
Keeping the above properties in mind, we now discuss how 
the plaquette and the dimer states compete with each other
when the inter-tetrahedron couplings $\lambda J$ and $\lambda J'$ are 
introduced. We expand the ground state energy 
up to the sixth order in $\lambda$ for several values of $j$.
We show the obtained energy  in Fig. \ref{fig:eg}, for which
the first-order inhomogeneous differential method \cite{Pade}
is applied to the bare series.
%%%%%%%%%%%%%%%%%%%%%%%%%%%%%%%%%%%%%%%%%%%%%%%%%%%%%%%%%%%%%%%%%
%%%%%%%%%%%%%%%%%%%%%%%%%%%%%%%%%%%%%%%%%%%%%%%%%%%%%%%%%%%%%%%%%
As mentioned above, for $\lambda\rightarrow0$, 
 the first-order quantum phase transition occurs between the 
plaquette and  the dimer phase at the critical point $j_c=1$.  
It is seen in the Fig. \ref{fig:eg} that  
the critical value for the phase transition is little changed 
even if we increase $\lambda$.
In fact, the energy up to the second order in $\lambda$
is same for both states near $j=1$,
as pointed out by Harris et al.\cite{Harris}  For small $\lambda$,
the first-order transition point is given by 
$j_c\sim1-0.021\lambda^3$.  Remarkably enough, 
 the first-order transition point is estimated 
as $j_c\sim1.0$ even for $\lambda=1$, where our generalized model
is reduced to the original one. 
Therefore, we arrive at a quite interesting conclusion that
the homogeneous spin system $(\lambda=j=1)$ with pyrochlore structure
is located quite closely to the phase boundary of the first-order
quantum phase transition, although  
it is difficult to definitely conclude which phase the ground state 
really belongs to within our accuracy. This fact clarifies the reason why 
Harris et al. and Isoda et al. had different conclusions 
on the nature of the 
ground state for the same $j=1$ model, where the former (latter)
claimed that the ground state is a dimer singlet (plaquette singlet).
As mentioned above, the energy for two phases is very close 
to each other, so that the mean-field type treatment may not 
correctly specify the ground state. Furthermore, 
there even remains the possibility that the system is 
just on the boundary, and thus the 
ground state could be 
degenerate at $j=1$. In any case, it is 
instructive to notice
that unusual dual-properties reflecting both natures of
the plaquette- and dimer-states should emerge around $j=1$ in various 
physical quantities such as the excitation spectrum, etc.

The results obtained above do not necessarily imply that 
a disordered ground state is always realized in the whole range of $j$.
It is needed to study how the disordered phases compete with 
possible antiferromagnetic phases driven by the three-dimensional (3D) 
exchange couplings.
We first recall that for $j=0$ and $\lambda=1$, 
the system should  have an antiferromagnetic order, 
as mentioned before.
On the other hand, in the case $j\rightarrow\infty (J'\rightarrow 0)$
the spin system is reduced to the $s=1/2$ massless Heisenberg chain 
characterized by the Tomonaga-Luttinger liquid phase.\cite{Bethe}
In the parameter regime ($j>1)$, a different type of the magnetic 
order was observed experimentally 
for $\rm GeCu_2O_4$ ($j\sim 6$).\cite{GeCu2O4}
Therefore,  we have to carefully check whether
the above different antiferromagnetic 
orders are indeed stabilized in our model.

\begin{figure}[htb]
\begin{center}
\includegraphics[width=8cm]{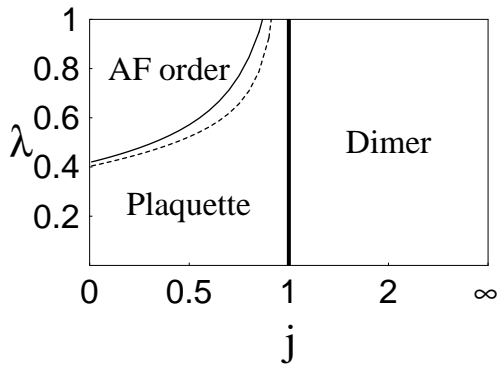}
\end{center}
\caption{Phase diagram for the $s=1/2$ quantum Heisenberg model with
pyrochlore structure. Bold line represents the phase boundary 
 which separates the dimer and the plaquette phases.
Solid (dashed) line indicates the phase boundary between
the plaquette and the magnetically ordered phases, which is
determined by the spin gap (staggered susceptibility).
%%%with  Dlog [1/2] Pad\'e approximants.
}
\label{fig:phase}
\end{figure}

We first study the magnetically ordered phase
in the region $0<j<1$.
To this end, we compute the susceptibility for a staggered field and 
the triplet excitation energy
up to the fourth order in $\lambda$ for various values of $j$.
To observe the second-order transition to the magnetically ordered phase,
we study the spin gap at ${\bf k}={\bf 0}$ in the Brillouin zone,
which should vanish at the phase transition point.
By applying Pad\'e approximants to computed series, \cite{Pade}
we obtain the phase boundaries shown in Fig. \ref{fig:phase}.
%%%%%%%%%%%%%%%%%%%%%%%%%%%%%%%%%%%%%%%%%%%%%%%%%%%%%%%%%%%%%%%%%
%%%%%%%%%%%%%%%%%%%%%%%%%%%%%%%%%%%%%%%%%%%%%%%%%%%%%%%%%%%%%%%%%
When $(j,\lambda )=(0,0)$, the system is reduced to an assembly of isolated
tetrahedra. With the increase of $\lambda$, the 3D network develops, enhancing
the antiferromagnetic correlation.
At last, the second-order quantum phase transition occurs to 
the magnetically ordered phase (e.g. the critical value 
is given by $\lambda_c\sim0.4$ for $j=0$).
The increase of $j$ suppresses the magnetic 
correlation due to strong frustration, and thus favors the plaquette 
phase. The critical value is estimated as $j_{c}\sim0.9$ for $\lambda=1$,
where our generalized model is reduced to the original system.
As seen in this figure, the boundaries determined in two
distinct ways slightly differ
from each other, which may be
due to the lower-order (fourth) series expansion done here.
Although it is desired to perform  a
higher-order calculation
to determine the phase boundary more precisely,
its essential feature is certainly given by the present calculation;
e.g. the magnetic phase is not  dominant for $j=1$, but its phase 
boundary is rather close to $j=1$.

We next examine another possibility of the antiferromagnetic order
observed for $\rm GeCu_2O_4$ \cite{GeCu2O4} in the region $j>1$. 
For this purpose, we calculate the susceptibility for 
the corresponding staggered field
up to the third order in $\lambda$.  As a result, we find that
the divergent singularity around $\lambda=1$ is gradually suppressed,
as $j$ is decreased from the value $(j=\infty)$ for 
the isolated spin chain.
This tendency implies that in the region $j>1$, 
the system does not enter the antiferromagnetically ordered phase, 
but always stays in the dimer phase with spin gap.
Therefore, it is seen from the above analysis that 
the magnetic order observed for the 
compound $\rm GeCu_2O_4$\cite{GeCu2O4} ($j \sim 6$)
may not be simply explained  in terms of the isotropic 
Heisenberg model employed here. This in turn suggests that 
 anisotropic interactions may be important 
to realize an ordered state in this compound.

%%%%%%%%%%%%%%%%%%%%%%%%%%%%%%%%%%%%%%%%
%%%%%%%%%%%%%%%%%%%%%%%%%%%%%%%%%%%%
%                   S=1 MODEL 
%%%%%%%%%%%%%%%%%%%%%%%%%%%%%%%%%%%%%%%%%
%%%%%%%%%%%%%%%%%%%%%%%%%%%%%%%%%%%

\begin{figure}[htb]
\begin{center}
\includegraphics[width=8cm]{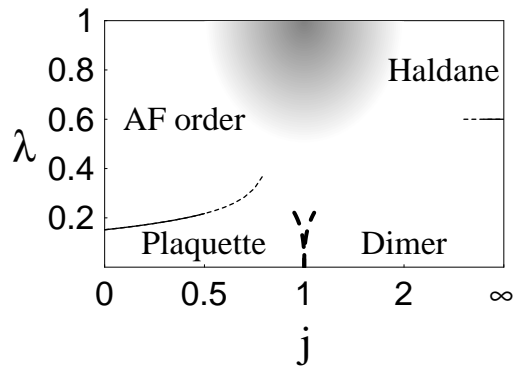}
\end{center}
\caption{Schematic phase diagram for the $s=1$ antiferromagnetic 
Heisenberg model with pyrochlore structure.}
\label{fig:s1phase}
\end{figure}

Let us now turn to the $s=1$ system 
on a pyrochlore lattice. \cite{Yamashita,mean-field}
%In the system with isotropic interactions, 
%the possibility of the chiral order due to the phonon 
%has been discussed by Yamashita and Ueda. \cite{Yamashita}
Although the series-expansion calculation
becomes much more difficult in this case,
we can still deduce some instructive comments on the $s=1$ pyrochlore lattice.
In order to use series expansion techniques,
we again start with an isolated $s=1$ tetrahedron,
whose eigenvalues are listed in Table \ref{II}.
%%%%%%%%%%%%%%%%%%%%%%%%%%%%%%%%%%%%%%%%%%%%%%%%%%%%%%%%%%%%%%%%%%%%%%%%%
\begin{table}[htb]
\caption{Eigenvalues of an isolated $s=1$ tetrahedron:
$E_{\rm D}$ ($E_{\rm P}$)
is the energy for the first (second) part in eq. (\ref{eq:iso}).
%%%the total energy is given by $E=E_{\rm D}+E_{\rm P}$.
}\label{II}
%\begin{ruledtabular}
\begin{tabular}{c|ccc|c|ccc|ccc|c|ccc|ccccc}
\toprule
$S_{12}$&\multicolumn{3}{c|}{0}&\multicolumn{7}{c|}{1}&\multicolumn{9}{c}{2}\\
$S_{34}$&0&1&2& 0&\multicolumn{3}{c|}{1}&\multicolumn{3}{c|}{2}& 
0&\multicolumn{3}{c|}{1}&\multicolumn{5}{c}{2}\\
$S_{\rm total}$&0&1&2& 0&0&1&2&1&2&3& 2&1&2&3&0&1&2&3&4\\
\hline
$E_{\rm D}/J$&-4&-3&-1& -3&\multicolumn{3}{c|}{-2}&\multicolumn{3}{c|}{0}&
-1&\multicolumn{3}{c|}{0}&\multicolumn{5}{c}{2}\\
$E_{\rm P}/J'$&0&0&0& 0&-2&-1&1&-3&-1&2& 0&-3&-1&2&-6&-5&-3&0&4\\
\botrule
\end{tabular}
%\end{ruledtabular}
\end{table}
%%%%%%%%%%%%%%%%%%%%%%%%%%%%%%%%%%%%%%%%%%%%%%%%%%%%%%%%%%%%%%%%%%%%%%%%%
It is seen that the $s=1$ plaquette-singlet 
with $S_1+S_2=S_3+S_4=2$ and $S_{\rm total}=0$ is the ground state for 
$0<j<1$ , whereas
  the dimer singlet
state with $S_1+S_2=S_3+S_4=0$ and $S_{\rm total}=0$ is the
ground state for $j>1$.  In contrast to the $s=1/2$ model, 
an isolated tetrahedron in the homogeneous point $(j=1)$
has three-fold degenerate ground states, 
which are composed of the above-mentioned  singlet states  together with
another singlet state with $S_1+S_2=S_3+S_4=1$ and $S_{\rm total}=0$,
which may be regarded as a $s=1/2$ plaquette-singlet 
state (see Table \ref{I}).
By turning on the inter-tetrahedron coupling, 
we observe how the above three-fold singlet states evolve  
 on a pyrochlore lattice.
To investigate the first-order quantum phase transitions among 
three phases, we estimate the ground state energy 
up to the fourth order in $\lambda$.
This expansion claims that in contrast to the 
$s=1/2$ case, there exists an intermediate 
($s=1/2$ plaquette) phase between 
the $s=1$ plaquette phase and the dimer phase in 
small $\lambda$: two phase boundaries are estimated as  
$j_c=1-0.42\lambda^3$ and $j_c=1+0.084\lambda^3$, which are 
shown as the bold dashed lines in Fig. \ref{fig:s1phase}.
%%%%%%%%%%%%%%%%%%%%%%%%%%%%%%%%%%%%%%%%%%%%%%%%%%%%%%%%%%%%%%%%%
%%%%%%%%%%%%%%%%%%%%%%%%%%%%%%%%%%%%%%%%%%%%%%%%%%%%%%%%%%%%%%%%%
Although we do not have a definite answer to the question 
whether this intermediate state
can survive as the ground state even for $\lambda=1$, 
it may be a possible candidate which competes with other states 
at fully frustrated point  $j=1$ with $\lambda=1$.
We next  calculate  the spin gap
 to see how stable the magnetically ordered phase for 
the $s=1$ case is in comparison with the $s=1/2$ case.
 By applying the 
%%%%Dlog [1/1] 
Pad\'e approximants to the third-order results, 
\cite{Pade} we deduce the phase boundary shown as the solid 
line in Fig. \ref{fig:s1phase}.
It is seen that the area of the magnetically ordered phase is 
more extensive than the $s=1/2$ case.
By recalling the phase diagram for the $s=1/2$ case, it is thus
expected that the magnetically ordered phase may be more dominant 
 around the homogeneous point $j=1$. Finally, we make a brief comment on  the 
 small $J'$ (large $j$) case.  For $J'=0$, the system is reduced to 
the $s=1$ spin chain with bond alternation,
where the dimer phase and the Haldane phase are separated
at the critical point $\lambda_c=0.6$.\cite{S1}
Although it is difficult to estimate the phase boundary
between these spin-gap states in the presence of
interchain coupling, it is naively expected
that the Haldane phase may disappear when $j$ is decreased 
down to $j=1$.

%%%%%%%%%%%%%%%%%%%%%%%%%%%%%%%%%%%%%%%%%%
%\section{Summary}
%%%%%%%%%%%%%%%%%%%%%%%%%%%%%%%%%%%%%%%%%%
In conclusion, we have discussed the ground-state phase diagram
for the  $s=1/2$ Heisenberg model 
with pyrochlore structure by means of the series expansion method.
 In particular, it has been found that
 the two different spin-gap states
strongly compete with each other around $j=1$, where the compound 
$\rm Y(Sc)Mn_2$\cite{Y(Sc)Mn2} may be located.  
Also, the antiferromagnetic
phase has been shown to be extended rather closely to the phase boundary.
Concerning $\rm GeCu_2O_4$\cite{GeCu2O4}, for which $j \sim 6$, 
it has turned out that 
the present model may not describe its magnetic order,
suggesting that some other mechanism
 should be considered for the magnetism.
For the  $s=1$ system, we have not been able to deduce the 
definite conclusion on the  phase diagram, but have checked that
the magnetically ordered phase may be more dominant
around  $j=1$ in comparison with  the $s=1/2$ case. Also,
besides the known states such as dimer, plaquette and
magnetically ordered states, another intermediate spin-gap state 
may also be a candidate for the ground state
around $j=1$.  Since this argument has been based on
the calculation for small $\lambda$,
it is desired to confirm whether this spin-gap state 
really takes part in the strong competition 
around $j=1$, which is now under consideration.

%%%%%%%%%%%%%%%%%%%%%%%%%%%%%%%%%%%%%%%%%%%%%%%%%%%%%%%%%%%%%%%
%\acknowledgments
We would like to thank K. Ueda, Y. Yamashita, K. Okunishi and
Y. Imai for useful discussions. 
The work is partly supported by a 
Grant-in-Aid from the Ministry of Education, Science, Sports, 
and Culture. 
A.K. is supported by the Japan Society 
for the Promotion of Science. 
%A part of numerical computations in this work was carried out 
%at the Yukawa Institute Computer Facility.

%%%%%%%%%%%%%%%%%%%%%%%%%%%%%%%%%%%%%%%%%%%%%%%%%%%%%%%%%%%%%%%%%%%%%
%                        REFERENCES                                 %
%%%%%%%%%%%%%%%%%%%%%%%%%%%%%%%%%%%%%%%%%%%%%%%%%%%%%%%%%%%%%%%%%%%%%
%

%%%

\end{document}